\documentclass[conference]{IEEEtran}
\usepackage[utf8]{inputenc}
\usepackage{times,graphics, dsfont, epsfig,amsmath,xspace,endnotes,pifont,multirow,rotating,listings,amssymb,algorithmic,color,caption,nicefrac,adjustbox,tabularx,mathtools, algorithmic,algorithm}
\usepackage{graphicx}
\usepackage{tabularx}

\newcolumntype{L}[1]{>{\raggedright\arraybackslash}p{#1}} 
\newcolumntype{C}[1]{>{\centering\arraybackslash}p{#1}} 
\newcolumntype{R}[1]{>{\raggedleft\arraybackslash}p{#1}} %
\usepackage{steinmetz}
\usepackage{color}
\usepackage[dvipsnames]{xcolor}
\usepackage{amsmath}
\usepackage{amssymb}
\usepackage{subcaption}
\usepackage{graphicx}
\usepackage[letterpaper, left=0.625in, right=0.625in, bottom=1in, top=0.75in]{geometry}

\usepackage{tikz}
\usepackage{amsthm}

\newcommand{\e} {{\bf{e}}}

\newcommand{\y} {{\bf{y}}}
\newcommand{\x} {{\bf{x}}}

\newcommand{\nariman}[1]{\textcolor{red}{#1}}
\newcommand{\amir}[1]{\textcolor{blue}{#1}}

\author{
\IEEEauthorblockN{Nariman Torkzaban}
\IEEEauthorblockA{\textit{University of Maryland, College Park}\\
College Park, MD\\
narimant@umd.edu}
\and
\IEEEauthorblockN{Mohammad Khojastepour}
\IEEEauthorblockA{\textit{NEC Laboratories,
America}\\
Princeton, NJ\\
amir@nec-labs.com}
\and
\IEEEauthorblockN{John S. Baras}
\IEEEauthorblockA{\textit{University of Maryland, College Park}\\
College Park, MD\\
baras@umd.edu}
}

\def\BibTeX{{\rm B\kern-.05em{\sc i\kern-.025em b}\kern-.08em
    T\kern-.1667em\lower.7ex\hbox{E}\kern-.125emX}}

\IEEEoverridecommandlockouts\IEEEpubid{\makebox[\columnwidth]{ 979-8-3503-1090-0/23/\$31.00 ©2023 IEEE \hfill} \hspace{\columnsep}\makebox[\columnwidth]{ }}

\begin{document}

\title{Blind Cyclic Prefix-based CFO Estimation in MIMO-OFDM Systems}



\maketitle
\begin{abstract}
Low-complexity estimation and correction of carrier frequency offset (CFO) are essential in orthogonal frequency division multiplexing (OFDM). In this paper, we propose a low-overhead blind CFO estimation technique based on cyclic prefix (CP), in multi-input multi-output (MIMO)-OFDM systems. We propose to use antenna diversity for CFO estimation. Given that the RF chains for all antenna elements at a communication node share the same clock, the carrier frequency offset (CFO) between two points may be estimated by using the combination of the received signal at all antennas. We improve our method by combining the antenna diversity with time diversity by considering the CP for multiple OFDM symbols. We provide a closed-form expression for CFO estimation and present algorithms that can considerably improve the CFO estimation performance at the expense of a linear increase in computational complexity. We validate the effectiveness of our estimation scheme via extensive numerical analysis. 
\end{abstract}



%
\IEEEpeerreviewmaketitle

\section{Introduction}

Orthogonal frequency-division Multiplexing (OFDM) has been widely adopted in wireless communications standards as a strong multi-carrier modulation technique due to its spectral efficiency and resistance against frequency selective fading. OFDM is considered an enabling technology for the fifth generation (5G) and beyond communications systems, complementing multi-input multi-output (MIMO). However, OFDM is vulnerable against carrier frequency offset (CFO) which is among the major impairments \cite{Mok19} between radio-frequency (RF) transceivers that are caused by the frequency mismatch between the local oscillators at the RF transceivers or by the Doppler shift. Such an impairment destroys the orthogonality among the subcarriers and results in severe performance degradation in multi-carrier systems. Therefore, it is essential to design CFO estimation and compensation techniques to avoid the decline in bit error rate (BER) at the receiver. 

To be more precise, CFO usually consists of two components when normalized to subcarrier spacing; (i) the \emph{integral} part that results in a circular shift in the indices of the subcarriers, resulting in frequency ambiguity, and (ii) the \emph{fractional} part that impacts the orthogonality of the subcarriers provoking inter-carrier interference (ICI). The integral CFO estimation has been studied in the literature \cite{Pha16}\cite{Mar18}\cite{Mar20}\cite{Roth21}. However, most of the efforts in the literature including the present work, have been focused on fractional CFO estimation and compensation. See \cite{Moh21} for a comprehensive survey on CFO estimation and compensation techniques and an illustrative example of how CFO is specified in wireless communications standards. 

CFO estimation approaches in the prior art can be mostly classified into two categories; (i) data-oriented, and (ii) non-data-oriented. The approaches of the first type, either employ time domain training symbol sequences \cite{Moo94}\cite{Sch97}\cite{Gho9} or rely on extensive use of frequency domain pilots\cite{Yu04}\cite{Lei04}\cite{Chen18}. In order to obtain high accuracy, such approaches will have to use a large number of training sequences in the time domain or occupy large bandwidth in the frequency domain, introducing an extra overhead and inevitably causing the performance degradation of MIMO OFDM systems. For this reason, the non-data-oriented (a. k. a. blind) CFO estimation techniques, have gained increasing attention over the past decade. 

A group of blind CFO estimators, utilize the null subcarriers (NS) in the OFDM blocks \cite{Gao07}\cite{Meng20}. The NS-based blind estimators exploit the fact that the ICI resulting from CFO will show up at the null subcarriers and can be used to estimate and correct the CFO parameters. NS-based methods usually formulate the CFO estimation as a polynomial optimization problem over multiple blocks of OFDM symbols and then employ ESPIRIT-like or MUSIC-like search algorithms, or polynomial rooting methods to solve the optimization problem. Given the complexity of the search methods and the need for multiple OFDM blocks for accurate estimation, the NS-based methods are of higher complexity. In \cite{Meng20}, the authors have introduced the novel concept of gap subcarriers (GS) and have exploited this concept to approximate the CFO estimation objective with a cosine function. Then the parameters of the cosine function are determined uniquely by observing the value of the cost function at three different trial CFO values. Then the CFO parameter can be easily estimated. This method skips the large overhead of the search methods and complex traditional polynomial rooting techniques.   

Another category of blind CFO estimators exploits the structure of the cyclic prefix (CP) to design low-complexity CFO estimators \cite{Lin16}\cite{Yang18}. the performance of the CP-based techniques may decline when the channel becomes more frequency-selective. In \cite{Lin16}, a CFO estimation technique relying on the remodulation of the received signal at the receiver is proposed in multipath environments. Specifically, the theoretical mean-squared error (MSE) for CFO estimation is presented as a closed-form solution. The authors carry out the Cramer-Rao Band (CRB) on the MSE of CFO estimation for multipath channels and based on these theoretical analyses propose a fine CFO estimation technique that is of low complexity. 



In this paper, we propose a low-complexity blind CFO estimation approach for a MIMO system in the uplink direction, where each mobile user (MU) may be served by one or multiple access points (APs). The main contribution of the paper are as follows. 

\begin{itemize}
    \item We propose to use antenna diversity for the purpose of CFO estimation. Given that the RF chains for all antenna elements at a communication node share the same clock, the CFO between two points may be estimated by using the combination of the  received signal at all antennas.  
    \item Our proposed scheme can also combine the antenna diversity with time diversity by considering the CP for multiple OFDM symbols.
    \item We provide a low-complexity closed-form expression and derive the Cramer-Rao lower bound for CFO estimation.
    \item We provide an algorithm that can considerably improve the CFO estimation performance at the expense of a linear increase in computational complexity.
\end{itemize}

The remainder of the paper is organized as follows. Section~\ref{sec:desc} describes the system model. In Section~\ref{sec:problem} we elaborate on our proposed method for CFO estimation. Section~\ref{sec:evaluation} presents our evaluation results, and finally, in Section~\ref{sec:conclusions}, we highlight our conclusions.

\section{System Model} 
\label{sec:desc}

We consider an OFDM system, where the AP and MU are employing  MIMO for which the antenna elements at each node operate according to a common local oscillator. Let $N$ denote the size of the Discrete Fourier Transform (DFT) and $L$ denote the size of the CP and $\tilde{N} = N+L$ is the total symbol length including the CP. The $k$-th time domain vector of the transmitted signal is given by
\begin{align}
\x_k = [ x[\tilde{N}k], x[\tilde{N}k+1], \ldots, x[\tilde{N}(k+1)-1)] ]^T   
\end{align}
where $x[i] = x[i+N]$, $\forall i, 1 \leq i \mod \tilde{N} \leq L$. The time series signal can be interpreted as blocks of length $\tilde{N}$ including the CP. We assume that the number of the channel taps does not exceed $L$ and hence the channel taps for the $m$-th antenna in the time domain may be represented by
$$h^{(m)}_0, h^{(m)}_1, \ldots, h^{(m)}_{L-1}.$$
Without considering the effect of CFO and noise, the received signal at antenna $m$ at time $i$ can be written as
\begin{align}
    r^{(m)}[i] = \sum_{l=0}^{L-1} x[i-l] h^{(m)}_l
\end{align}
which depends on the current and the past $L-1$ time domain transmitted signals due to the effect of the multipath channel.
Let $\theta_k= N \Delta f / f_s$ denote the normalized CFO for the $k$-th OFDM symbol with respect to the first received time domain signal in the time frame $k$ where the CFO is $\Delta f$ in Hz and $f_s$ is the sampling frequency. We only consider fractional CFO, i.e., $-0.5 \leq \theta_k \leq 0.5$. We note that the normalized CFO in a multi-user scenario where each OFDM symbol may be transmitted from a different user may be different for different received time domain vectors. Nonetheless, if two adjacent symbols $k$ and $k+1$ belong to the same user $\theta_k = \theta_{k+1}$, but the first received time domain signal $x[(k+1)\tilde{N}]$ for the received vector $k+1$ is rotated by $\e^{j 2 \pi \frac{k\tilde{N}}{N} \theta_k}$ with respect to the first received time domain signal $x[k\tilde{N}]$ of the received vector $k$.
Let $\psi[i]$ denote the CFO for the $i$-th received signal which can be written as $\psi[k\tilde{N}+i] = e^{j 2 \pi \theta_k i /N}$ with respect to the first symbol of the $\x_k$. By considering the effect of CFO and noise, the received signal at antenna $m$ at time $i$ may be written as
\begin{align}\label{eq:rec_signal}
    r^{(m)}[i] = \sum_{l=0}^{L-1} \psi[i-l] x[i-l] h^{(m)}_l + z^{(m)}[i]
\end{align}
where $z[i]$ is the AWGN noise for the received signal at time $i$ with the variance of $\sigma_z^2$.

\section{Proposed CFO Estimation Technique}
\label{sec:problem}
Consider the $k$-th OFDM symbol received at the $m$-th antenna element. Let $\xi$ be a variable that will later be useful for CFO estimation and define the vector $\y^{(m)}_k(\xi)$ of length $2L$ with $\ell$-th element given by, 
\begin{align}
    y^{(m)}_k[\ell](\xi) = \left(r^{(m)}[k\tilde{N}+l]- e^{j 2 \pi \xi} r^{(m)}[k\tilde{N}+N+l]\right)
\end{align}

The auto-correlation function of the vector $\y^{(m)}_k(\xi)$ in the expanded form is given by, 
\begin{align} \label{eq:obj}
    J_{k,m}(\xi) =& 
    \mathbb{E}\left\{\sum_{l=0}^{2L-1} y^{(m)}_k[\ell](\xi)y^{(m)}_k[\ell](\xi)^*\right\}
\end{align}
where $\rho = e^{j 2 \pi \xi}$, and $(.)^*$ is the Hermitian operator. 
We note that due to the definition of CP and the fact that the transmitted signal is random with zero mean, for $0 \leq i,j \leq N+L-1$, we have 
\begin{align}\label{eq:aux_delta}
    \mathbb{E}\{ x[k\tilde{N}+i] x[k\tilde{N}+j]\} = \sigma_x^2\delta(|i-j| \mod N)
\end{align}
%
where $\delta(.)$ is the Kronecker delta function and $\sigma_x^2$ is the power of the time domain signal per sample. Using equations~\eqref{eq:rec_signal} and~\eqref{eq:aux_delta}, after straightforward algebraic operations followed by the expansion of \eqref{eq:obj}, $J_{k,m}(\xi)$ is simplified as
\begin{align}
    J_{k,m}(\xi) = 2L \eta^{(m)} (1-cos(2 \pi (\xi - \theta_k)) + 2 L \sigma_z^2\label{eq: cost}
\end{align}
where $\eta^{(m)} = \sigma_x^2 \sum_{l=0}^{L-1} |h^{(m)}_l|^2$ is independent of $\xi$.
%
The simplified form of the auto-correlation function in \eqref{eq: cost} reveals that the function achieves its minimum with respect to $\xi$ at $\xi = \theta_k$. Therefore, it makes sense to minimize equation~\eqref{eq:obj} as a cost function to obtain $\xi$ as the estimate for the CFO parameter $\theta_k$. If the receiver is equipped with $M$ antennas, the cost function can be redefined as the summation over all antennas  
\begin{align}\label{eq:cost_M}
    J_k^M(\xi) &= \sum_{m=1}^{M} J_{k,m}(\xi) \nonumber \\
    &= 2L \left(\sum_{m=1}^{M} \eta^{(m)} \right) (1-cos(2 \pi (\xi - \theta_k)) + 2 L M \sigma_z^2.
\end{align}

Clearly, the cost function $J_k^M(\xi)$ for multiple antennae also has its minimum at $\xi = \theta_k$. 

\subsection{Coarse Estimation}

For large values of M, the cost function \eqref{eq:cost_M} is empirically given by, 
\begin{align}\label{eq:cost_M_est}
    J^M_k(\xi) \approx& \sum_{m=1}^{M}  \sum_{l=0}^{2L-1} \left(y^{(m)}_k[\ell](\xi)\right) \left(y^{(m)}_k[\ell](\xi)\right)^*. 
\end{align}

The minimizer of \eqref{eq:cost_M_est} is found by setting its derivative with respect to $\xi$ to zero which would uniquely provide the CFO estimate as, 
\begin{align}
    \xi = \frac{1}{2 \pi} \angle \sum_{m=1}^{M}  \sum_{l=0}^{2L-1} r^{(m)}[k\tilde{N}+l] r^{(m)}[k\tilde{N}+N+l]^* 
\end{align}

If multiple OFDM symbols are received from a single user (without loss of generality, denoted by symbols $k$, $k = 1, \ldots, K$), the cost function can be better approximated as
\begin{align}\label{eq:obj_MK}
    J^{K,M}(\xi) &\approx\nonumber \\ & \sum_{k=1}^{K} \sum_{m=1}^{M}  \sum_{l=0}^{2L-1} \left(y^{(m)}_k[\ell](\xi)\right)\left(y^{(m)}_k[\ell](\xi)\right)^* 
\end{align}
and the CFO estimate would be found as, 
\begin{align} \label{eq:zeta}
    \xi = \frac{1}{2 \pi} \angle \sum_{k=1}^{K} \sum_{m=1}^{M}  \sum_{l=0}^{2L-1} r^{(m)}[k\tilde{N}+l] 
    r^{(m)}[k\tilde{N}+N+l]^* 
\end{align}

The above solution for $M=1$ antenna is similar to the solution found in \cite{Lin16} for coarse CFO estimation using $K$ OFDM symbols. The solution given by \eqref{eq:zeta} relies on the independence between the time domain transmitted symbols as well as the independence between the channel coefficients. The larger the ensemble over which the summation is calculated, the better the approximation. Hence, increasing the number of OFDM symbols $K$ or the number of antennas $M$ would increase the estimation performance.

\subsection{Fine Estimation}

We note that for each $k$, $1 \leq k \leq K$ and $m$, $1 \leq m \leq M$ the objective function is comprised of $2L$ terms which contribute to the objective function of the form
\begin{align}
    J_{k,m}(\xi) = A - B cos(2 \pi (\xi - \theta_k)
\end{align}

We note that for such a convex objective function, the larger the second derivative the lower the estimation error of $\xi$. The objective function in \eqref{eq:obj_MK} can be written as
\begin{align}\label{eq:obj_MK_sum}
    J^{K,M}(\xi)  &= \nonumber\\&
    \sum_{l=0}^{2L-1} \mathbb{E} \left\{ \left(y^{(m)}_k[\ell](\xi)\right) \right.\left.
     \left(y^{(m)}_k[\ell](\xi)\right)^* \right\}  =\sum_{l=0}^{2L-1} R(l)
\end{align}
and therefore, the second derivative of \eqref{eq:obj_MK} is given by the summation of the second derivatives of $R(l)$, i.e., 
\begin{align}
    \frac{\partial^2}{\partial \xi^2} J^{K,M}(\xi)  =  \sum_{l=0}^{2L-1}  \frac{\partial^2}{\partial \xi^2} R(l)
\end{align}
Note that, $R(l)$ can be stated in the summation form as 
\begin{align}
    R(l) &= \frac{1}{M} \sum_{m=1}^M R^{(m)}(l)
\end{align}
where,
\begin{align}\label{eq:RL}
    R^{(m)}(l) = 2(\eta^{(m)} - \eta^{(m)}_l \: cos(2 \pi (\xi - \theta_k))) - 2 \sigma_z^2
\end{align}
with,
\[ \begin{cases} 
    \eta^{(m)}_l = \sigma_x^2 \sum_{i=0}^{l} |h^{(m)}_i|^2
& 0 \leq i \leq L-1 \\
      \eta^{(m)}_l = \sigma_x^2 \sum_{i=l-l}^{L-1} |h^{(m)}_i|^2 & L \leq i \leq 2L-1 
   \end{cases}
\]
Hence, $R(l)$ is obtained as
\begin{align}
    R(l) = 2 \sum_{m=1}^M \eta^{(m)} - 2 \sum_{m=1}^M \eta^{(m)}_l cos(2 \pi (\xi - \theta_k)) + 2 \sigma_z^2
\end{align}
and its second derivative with respect to $\xi$ is given by
\begin{align}\label{eq:RLD}
    \frac{\partial^2}{\partial \xi^2} R^{(m)}(l) = 8\pi^2 cos(2 \pi (\xi - \theta_k))) \sum_{m=1}^M \eta^{(m)}_l 
\end{align}
Comparing \eqref{eq:RL} and \eqref{eq:RLD}, it is noted that the smaller the value of $R(l)$ the larger its derivative. Hence, one can improve the estimation performance by using a summation over a proper subset of indices, i.e., $ \{0,1, \ldots, 2L-1\}$, that contribute to the objective function \eqref{eq:obj_MK_sum}. Let us denote this subset by $S(\Lambda)$. The parameter $\Lambda$ is the size of the set $S(\Lambda)$ which is important to be chosen such that the estimation error is minimized. For a given index subset $S(\Lambda)$, the objective function is modified as
\begin{align}\label{eq:obj_MK_subset}
    J^{K,M}(\xi) &=\nonumber\\ &\sum_{l \in S(\Lambda)} R(l) =
    \sum_{l \in S(\Lambda)} \mathbb{E} \left\{ \left(y^{(m)}_k[\ell](\xi)\right)
     \left(y^{(m)}_k[\ell](\xi)\right)^* \right\} 
\end{align}
where $R(l)$ can be empirically found as
\begin{align}\label{eq:r_l_compute}
    R(l) \approx & \sum_{k=1}^{K} \sum_{m=1}^{M}  \left(y^{(m)}_k[\ell](\xi)\right) \left(y^{(m)}_k[\ell](\xi)\right)^* 
\end{align}
The fine estimate is then given by
\begin{align} \label{eq:zeta_fine}
    \xi = \frac{1}{2 \pi} \angle \sum_{k=1}^{K} \sum_{m=1}^{M}  \sum_{l \in S(\Lambda)} r^{(m)}[k\tilde{N}+l] 
    r^{(m)}[k\tilde{N}+N+l]^* 
\end{align}

Algorithm~\ref{alg:f_fine} provides the steps required to perform a fine CFO estimation for a pre-selected value of $\Lambda$. This algorithm may be run for a fixed value of $\Lambda$. For example $\Lambda = L$ provides a good estimation as will be shown later in the evaluation section.
However, the algorithm can be further improved by finding the value of $\Lambda$ adaptively. The detailed of adaptive-fine CFO estimation is given in Algorithm~\ref{alg:a_fine}. 
This algorithm provides significant improvement with only two iterations. Please refer to the results in the evaluation section.

\begin{algorithm}
\caption{\textit{Fixed-Fine-Estimate}}
\label{alg:f_fine}
 \begin{algorithmic}[1]
 \renewcommand{\algorithmicrequire}{\textbf{Input:}}
 \renewcommand{\algorithmicensure}{\textbf{Output:}}
 \REQUIRE $\Lambda$, $ 1 \leq \Lambda \leq 2L-1$
 \\ 
 \STATE Compute the coarse estimate of $\theta_k$ using \eqref{eq:zeta}.
 \STATE Calculate $R(l)$, $0 \leq l \leq 2L-1$ using the approximation in \eqref{eq:r_l_compute}.
 \STATE find the set $S(\Lambda)$ which contains the indices of $\Lambda$ smallest values of $R(l)$.
\STATE $\hat{\theta}\xleftarrow{}$ Calculate the fine estimate via equation \eqref{eq:zeta_fine}.
\STATE \textbf{Return} $\hat{\theta}$.
\end{algorithmic}
\end{algorithm}
\begin{algorithm}
\caption{\textit{Adaptive-Fine-Estimate}}
\label{alg:a_fine}
 \begin{algorithmic}[1]
 \renewcommand{\algorithmicrequire}{\textbf{Input:}}
 \renewcommand{\algorithmicensure}{\textbf{Output:}}
 \REQUIRE $ITER$
 \STATE $iter =0$ 
 \STATE $\hat{\theta}\xleftarrow{}$ the coarse estimate of $\theta_k$ using \eqref{eq:zeta}.
 \REPEAT{
 \STATE Calculate $R(l)$, $0 \leq l \leq 2L-1$ using the approximation in \eqref{eq:r_l_compute}
 \STATE \textbf{For} $\Lambda = 1:1:2L-1$:
 \STATE \quad \quad Compute $S(\Lambda)$ which contains the indices of $\Lambda$ \\
  \quad \quad smallest values of $R(l)$. 
 \STATE \quad \quad $\phi({\Lambda})\xleftarrow{}$ Calculate the fine estimate via \eqref{eq:zeta_fine}.
 \STATE \textbf{End For}
 \STATE $\Lambda^{\mathrm{OPT}}\xleftarrow{}$ $\arg\min_{ 1 \leq \Lambda \leq 2L-1} |\hat{\theta}-\phi({\Lambda})|$
 \STATE $\hat{\theta}\xleftarrow{} \phi({\Lambda^{\mathrm{OPT}}})$
 \STATE $iter ++$
 }
 \UNTIL{$iter = ITER$}
\STATE \textbf{Return} $\hat{\theta}$
\end{algorithmic}
\end{algorithm}

\section{Performance Evaluation}
\label{sec:evaluation}

\begin{figure*}[t]
\centering
\begin{subfigure}{0.49\textwidth}
    \includegraphics[width=\textwidth]{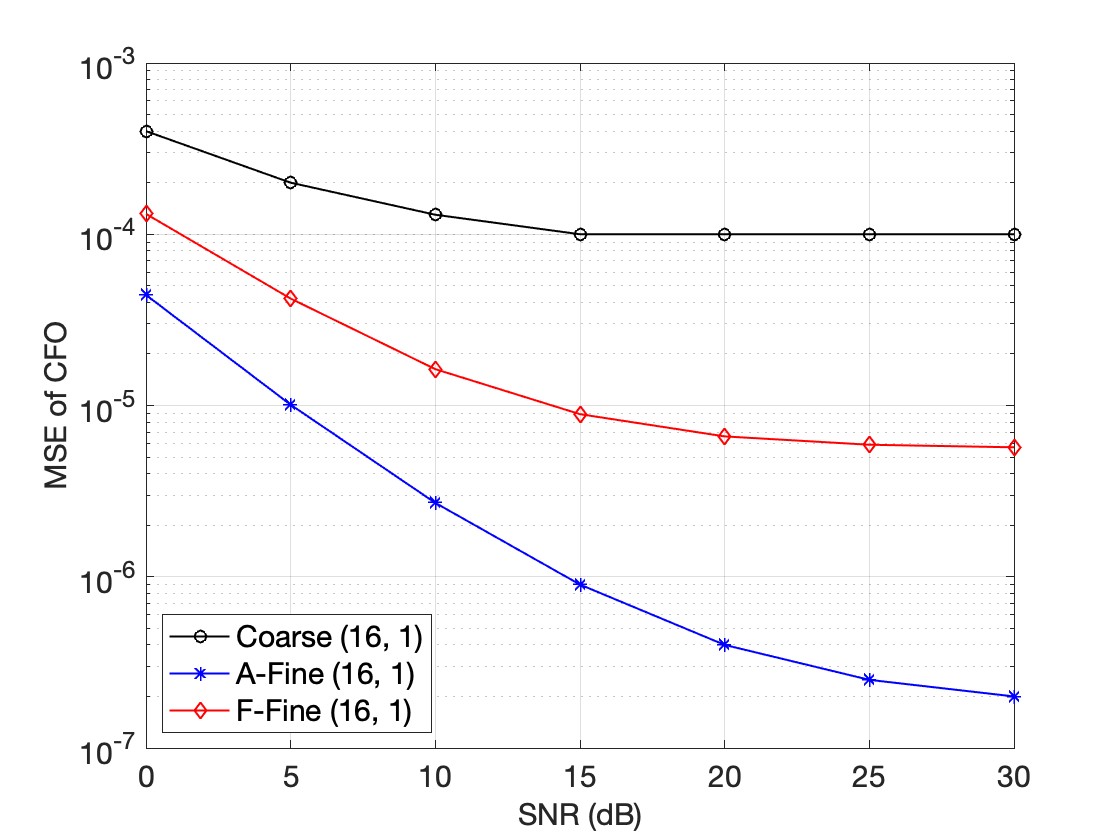}
    \caption{CFO estimation with $(K,M) = (16,1)$}
    \label{fig:exp1_Lin8}
\end{subfigure}
\hfill
\begin{subfigure}{0.49\textwidth}
    \includegraphics[width=\textwidth]{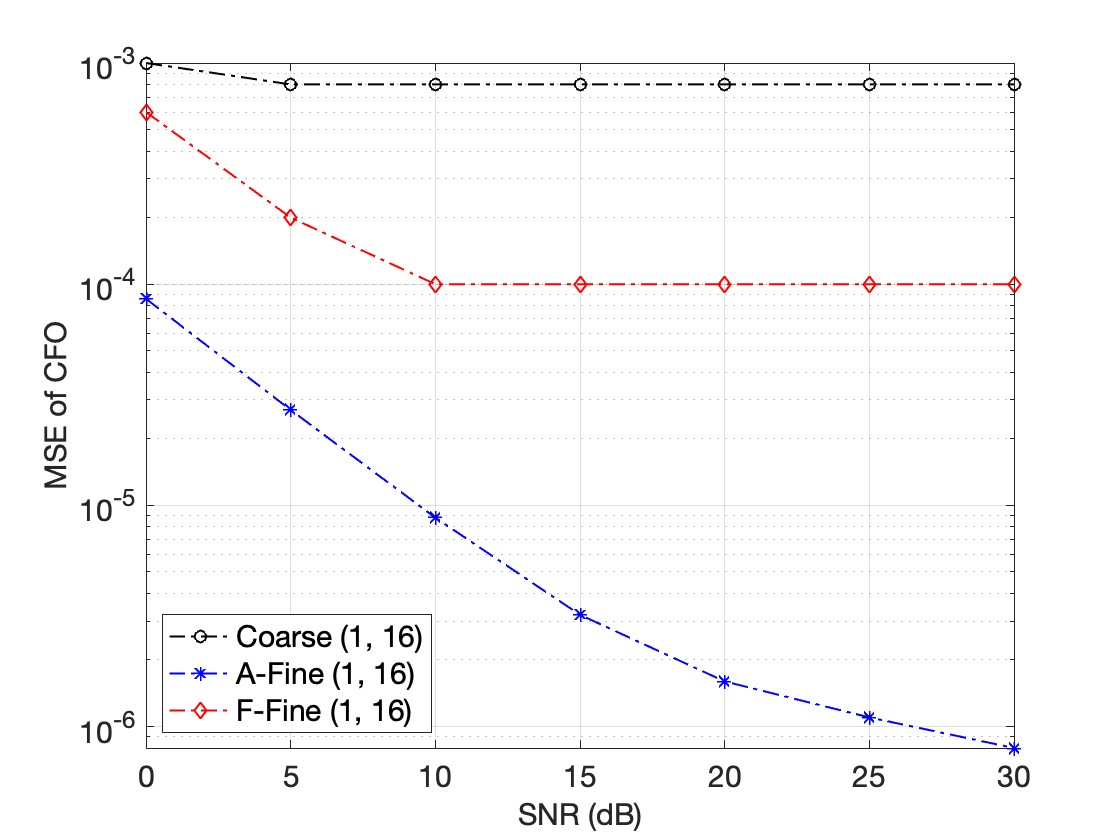}
    \caption{CFO Estimation with $(K,M) = (1,16)$}
    \label{fig:exp1_Khoj8}
\end{subfigure}
\caption{Comparison of coarse and fine CFO estimation with $M$ antenna elements and $K$ OFDM symbols }
\label{fig:exp1}
\end{figure*}

\begin{figure}
    \includegraphics[width=\linewidth]{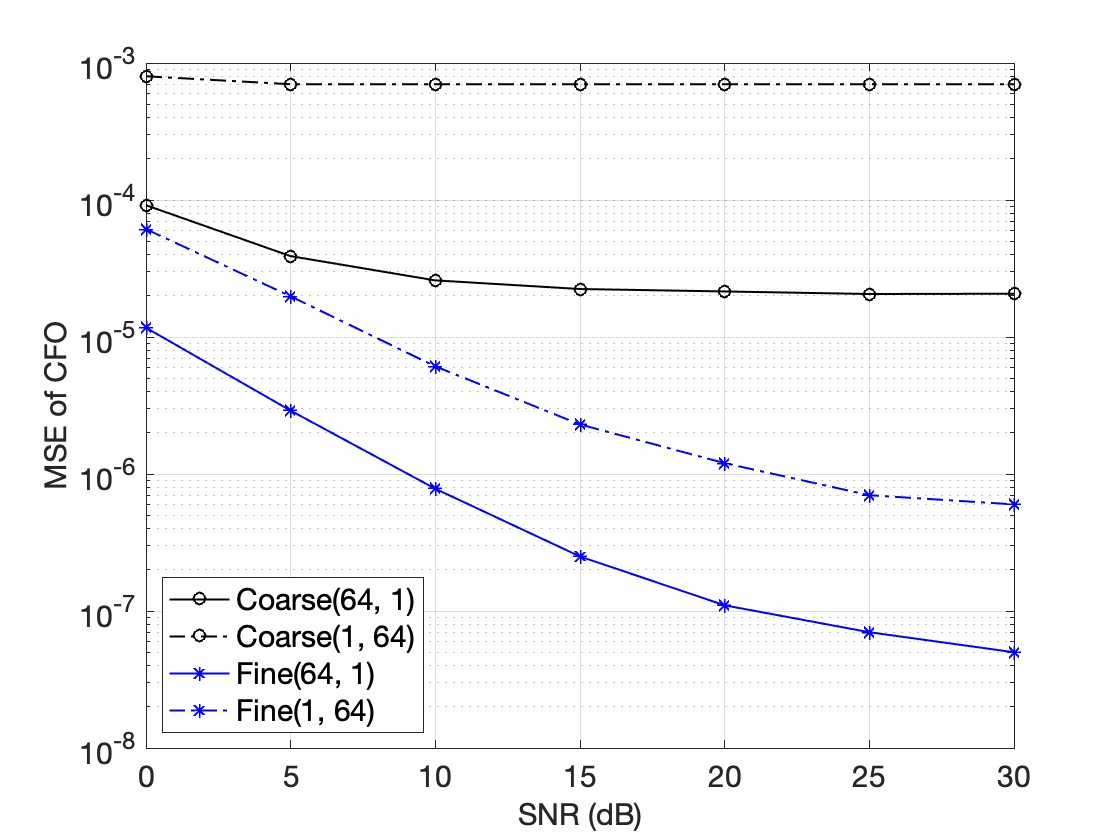}
    \caption{ Time vs. Antenna Diversity $(K,M)=(64,1)$, $(1,64)$}
    \label{fig:exp2_64}
\end{figure}

\begin{figure*}
\centering
\begin{minipage}{0.48\textwidth}
    \includegraphics[width=\textwidth]{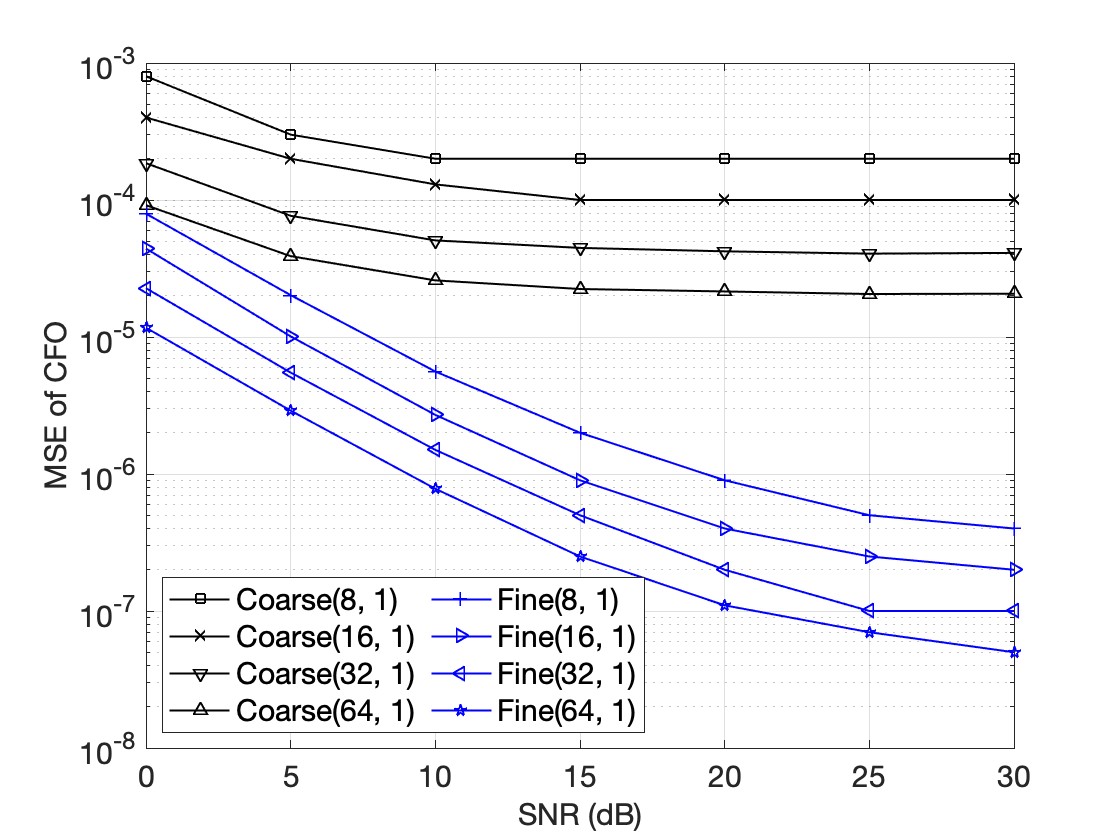}
    \caption{Impact of $K$, for $M=1$}
    \label{fig:exp3_a}
\end{minipage}
\begin{minipage}{0.48\textwidth}
    \includegraphics[width=\textwidth]{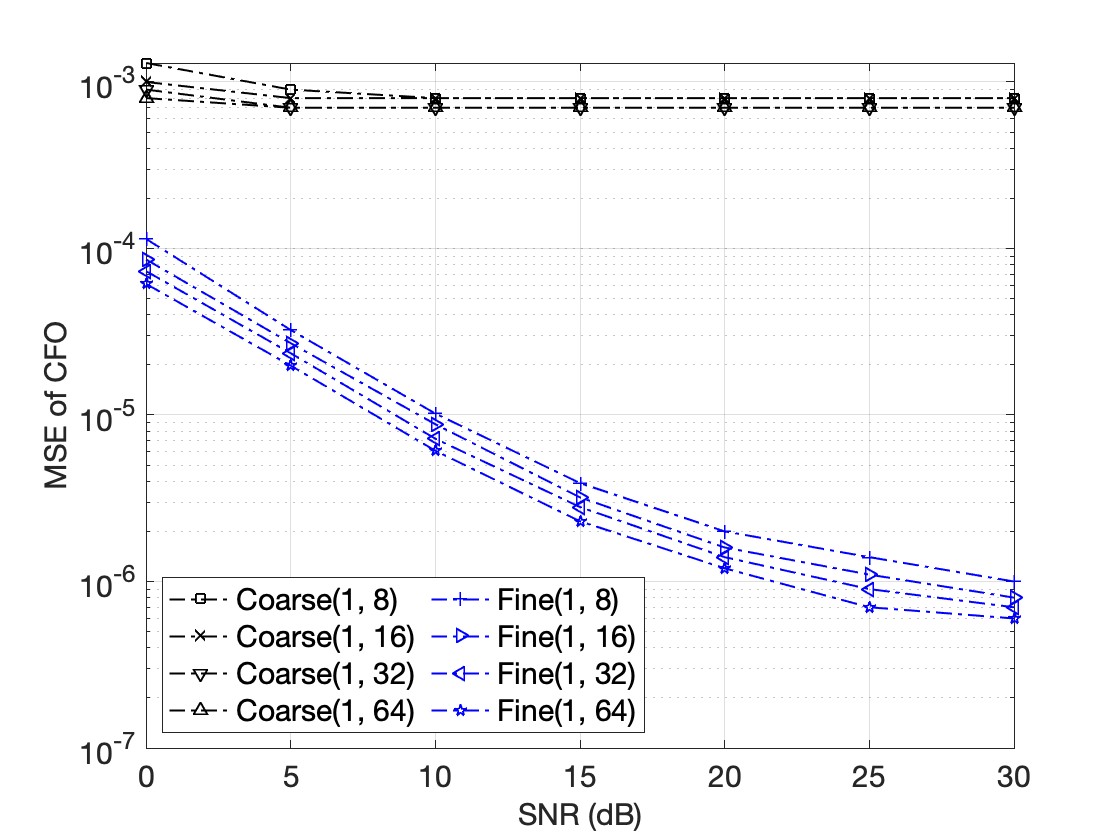}
    \caption{Impact of $M$, for $K=1$}
    \label{fig:exp3_b}
\end{minipage}
\end{figure*}

\begin{figure*}[t]
\centering
\begin{subfigure}{0.49\textwidth}
    \includegraphics[width=\textwidth]{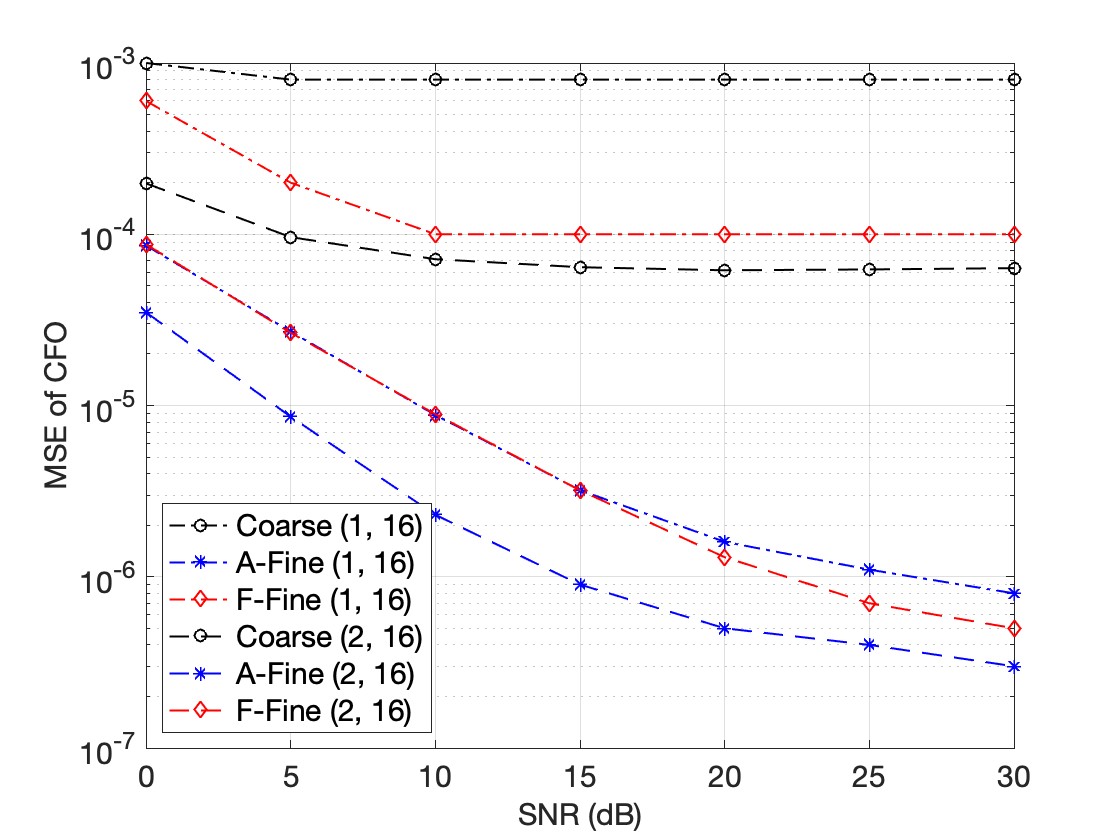}
    \caption{Comparison of coarse and fine CFO estimation}
    \label{fig:exp4_a}
\end{subfigure}
\hfill
\begin{subfigure}{0.49\textwidth}
    \includegraphics[width=\textwidth]{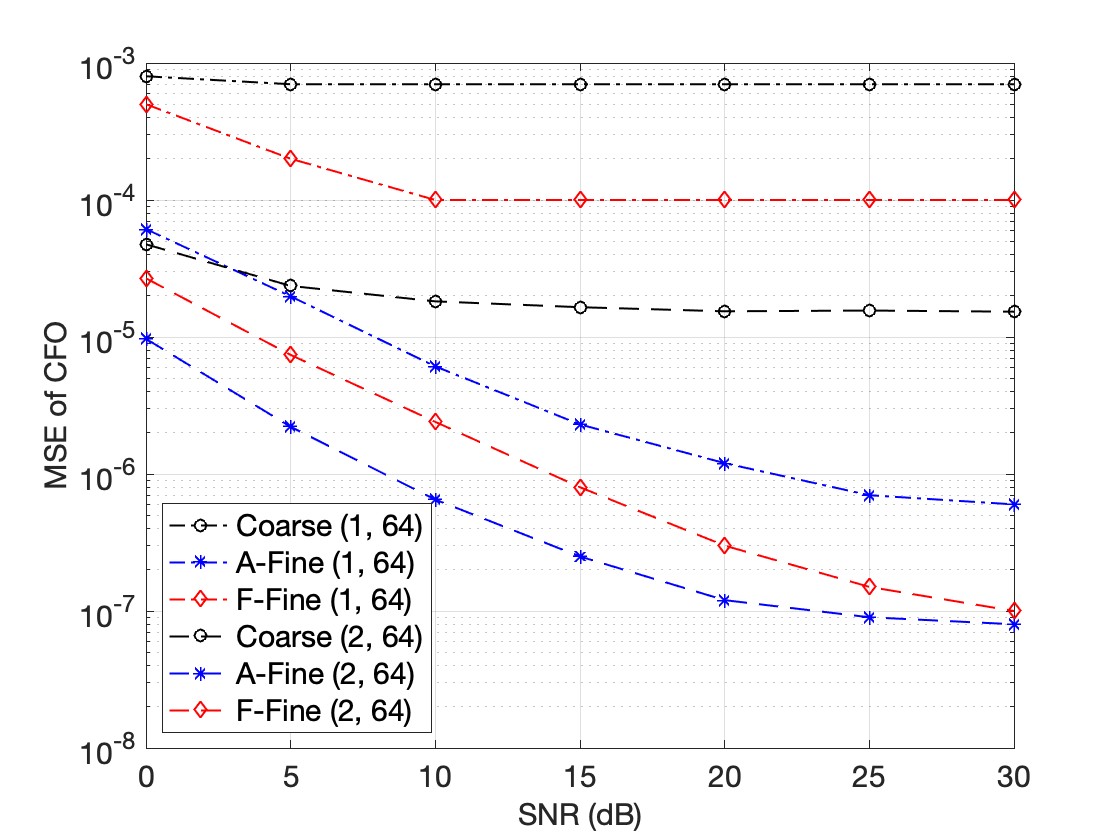}
    \caption{Comparison of coarse and fine CFO estimation.}
    \label{fig:exp4_b}
\end{subfigure}
\caption{Evaluation of combining time and antenna diversity for various $K$,  and $M$}
\label{fig:exp4}
\end{figure*}

In this section, we evaluate the performance of our CFO estimation technique under various circumstances. 

\subsection{Simulation Setup and Parameters}
Without loss of generality, we assume the MUs have only a single antenna embedded, while each AP may employ multiple antenna elements. We consider the transmission of symbols from a $16$-QAM constellation. Throughout the experiments, we set the DFT size $N=64$, CP length $L=16$, and the number of antenna elements per AP $M=1, 8, 16, 32, 64$. We adopt a multipath Rayleigh fading channel with $T=5$ taps with adaptive white Gaussian noise (AWGN), and the CFO value is set to $0.295$. Our tests are carried out by MATLAB on a server with an Intel i9 CPU at 2.3 GHz and 16 GBs of main memory and each simulation is run for $I= 10000$ trials and the results are averaged. We use the following metrics to evaluate the performance of our CFO estimation method: 

\begin{itemize}
    \item \textbf{Estimation Mode} may be \emph{coarse} or \emph{fine}. Fine estimation can be implemented in two \emph{adaptive} and \emph{fixed} modes based on the static or dynamic choice of the parameter $\Lambda$.
    \item \textbf{Time Diversity} corresponds to when CFO estimation is carried out across time; i. e. over multiple OFDM symbols. 
    \item \textbf{Antenna Diversity} corresponds to the case where CFO estimation is carried out across multiple antenna elements employed on a single AP. 
\end{itemize}

\subsection{Numerical Results}


Fig.~\ref{fig:exp1} depicts the MSE of CFO estimation under different modes, i. e. coarse, adaptive fine (a-fine), and fixed fine (f-fine). Fig.~\ref{fig:exp1_Lin8} corresponds to the case of time diversity with $K=16$, i.e. where a single-antenna MU communicates $16$ OFDM symbols to a single-antenna AP. While, Fig.~\ref{fig:exp1_Khoj8} corresponds to the case of antenna diversity with $M=16$, i. e. where a single-antenna MU communicates a single OFDM symbol to $16$ antenna elements of the AP. The value of $\Lambda = 16$ is set for the fixed fine CFO estimation. It is observed that the performance under adaptive fine estimation is superior to the coarse estimation by at least two orders of magnitude. Also, both the coarse estimation and fixed-fine estimation techniques saturate at some MSE value while the adaptive fine estimation method continues it descending trend even after $30 dB$ SNR.

Fig.~\ref{fig:exp2_64} compares the performance of CFO estimation under time diversity and antenna diversity, for $K=M=64$. It is observed that the MSE plot for CFO estimation under antenna diversity will lie and saturate above the MSE plot under time diversity, for coarse and fine estimation, respectively. The reason for this observation is that, although the channel tap coefficients for the antenna elements at a single AP are uncorrelated, they still follow the same power profile. Therefore, approximating the expectation with a summation needs larger values of $M$ in antenna diversity than $K$ in time diversity. 


Fig.~\ref{fig:exp3_a} and~\ref{fig:exp3_b} illustrate the effect of increasing the value of parameters $K$ and $M$ on the performance of CFO estimation, under time diversity and antenna diversity, respectively.  It is observed that increasing the value of parameters $K$ and $M$ shifts the CFO MSE plots downwards and improves the performance of the CFO estimation. Further, it is observed that increasing $K$ from $8$ to $64$ under time diversity improves the performance of CFO estimation by an order of magnitude while increasing $M$ has a less improving effect under the antenna diversity scheme.

Fig.~\ref{fig:exp4} shows how combining antenna diversity and time diversity improves the performance of CFO estimation. Fig.~\ref{fig:exp4_a}, and~\ref{fig:exp4_b} show this effect for $M=16$ and $M=64$, respectively. Interestingly, we observe that increasing the OFDM symbols results in more improvement in the case of $M=64$ compared to the case of $M=16$.

\section{Conclusions}
\label{sec:conclusions}

This paper presented a low-complexity CP-based blind CFO estimation for MIMO-OFDM systems. We used antenna diversity for the purpose of CFO estimation. Given that the RF chains for all antenna elements at a communication node share the same clock, the carrier frequency offset (CFO) between two points may be estimated by using the combination of the  received signal at all antennas. Further, we incorporated the notion of time diversity into the CFO estimation by considering the CP for multiple consecutive OFDM blocks. We defined a cost function employing the correlation of the received OFDM signals at multiple antenna and multiple OFDM blocks. We proposed algorithms for estimating the CFO with low complexity. Numerical results verify the validity of our proposed CFO estimation technique. 



\renewcommand{\nariman}[1]{\textcolor{red}{#1}}
\renewcommand{\amir}[1]{\textcolor{blue}{#1}}



%

\bibliographystyle{IEEEtran}
\bibliography{bibliography}

\end{document}